\documentclass[%
 reprint,
 amsmath,amssymb,
 aps,
]{revtex4-1}

\usepackage{graphicx}
\usepackage{dcolumn}
\usepackage{bm}
\usepackage{amsmath}
\usepackage{enumerate}
\usepackage{setspace}
\usepackage{dsfont}
\usepackage[pdfstartview=FitH,
            CJKbookmarks=true,
            bookmarksnumbered=true,
            bookmarksopen=true,
            colorlinks, 
            pdfborder=001,   
            linkcolor=blue,
            anchorcolor=blue,
            citecolor=blue
            ]{hyperref}

\begin{document}

\title{Improving parameter estimation of entropic uncertainty relation in continuous-variable quantum key distribution}
\author{Ziyang Chen$^{1}$}
\author{Yichen Zhang$^{2}$}
\author{Xiangyu Wang$^2$}
\author{Song Yu$^{2}$}
\author{Hong Guo$^{1}$}
\thanks{hongguo@pku.edu.cn}
\affiliation{$^1$ State Key Laboratory of Advanced Optical Communication, Systems and Networks, Department of Electronics, and Center for Quantum Information Technology, Peking University, Beijing 100871, China}
\affiliation{$^2$ State Key Laboratory of Information Photonics and Optical Communications, Beijing University of Posts and Telecommunications, Beijing 100876, China}


\date{\today}

\begin{abstract}
The entropic uncertainty relation (EUR) is of significant importance in the security proof of continuous-variable quantum key distribution under coherent attacks. The parameter estimation in the EUR method contains the estimation of the covariance matrix (CM), as well as the max-entropy. The discussions in previous works have not involved the effect of finite-size on estimating the CM, which will further affect the estimation of leakage information. In this work, we address this issue by adapting the parameter estimation technique to the EUR analysis method under composable security frameworks. We also use the double-data modulation method to improve the parameter estimation step, where all the states can be exploited for both parameter estimation and key generation; thus, the statistical fluctuation of estimating the max-entropy disappears. The result shows that the adapted method can effectively estimate parameters in EUR analysis. Moreover, the double-data modulation method can, to a large extent, save the key consumption, which further improves the performance in practical implementations of the EUR.
\end{abstract}

\pacs{03.67.Dd, 03.67.Hk}
\maketitle


\section{Introduction}
The quantum key distribution (QKD)~\cite{Gisin_RMP_2002, Scarani_RMP_2007, Weedbrook_RMP_2012, Diamanti_entropy_2015, Pirandola.arXiv.1906.01645} is one of the most mature quantum cryptography technologies, which can provide information-theoretical provable security together with the one-time pad method. The idea of QKD is to employ the basic principles of quantum physics to ensure the security of random keys and to use classical post-processing methods to find potential eavesdropping behaviors. Based on the dimension of the Hilbert space of the encoding, QKD can be roughly divided into two categories. One kind of protocol is
called the discrete-variable (DV) protocol, in which the dimension of the Hilbert space is finite. DV-QKD protocols have the superiority of long transmission distance, but depending on high-performance dedicated devices such as single-photon detectors. As an alternative, continuous-variable (CV) protocols, which use the infinite dimension of Hilbert space as the key space, give us opportunities to achieve the QKD process via off-the-shelf commercial components, e.g., homodyne detector and heterodyne detector.

The first idea of the CV-QKD protocol was exploiting squeezed states to carry the key information~\cite{Ralph_PRA_1999, Hillery_PRA_2000, Cerf_PRA_2001, Usenko_NJP_2011}. Then, in order to weaken the dependence on the squeezed-state sources, the coherent-state-based CV-QKD protocols were proposed~\cite{Grosshans_PRL_2002, Grosshans_nature_2003, Weedbrook_PRL_2004}. During these twenty years, research on protocol design and corresponding experimental verification was developing rapidly. Different novel CV-QKD protocols have been proposed, such as the two-way protocol~\cite{Pirandola_NatPhys_2008, Sun_Int.J.Quantum.Inf_2012, Zhang_J.Phys.B_2014, Ottaviani_PRA_2015, Ottaviani_Sci.Rep_2016, Zhang_JPB_2017}, the discrete modulation protocol~\cite{Leverrier_PRL_2009, Leverrier_PRA_2011, Li_arXiv_2018}, the measurement-device-independent (MDI) protocol~\cite{Li_PRA_2014, Zhang_PRA_2014, Pirandola_Nat.Photon_2015, Phys.Rev.A.96.042334.2017, Phys.Rev.A.96.042332.2017, Phys.Rev.A.97.052327.2018, Chen_PRA_2018}, etc., each of which has its own advantages in different scenarios. Besides the protocol design, the experiments also have made a tremendous step forward with the progress of today's technology~\cite{Jouguet_OE_2012, Jouguet_Nat.Photon_2013, Zhang_arXiv_2017}.

The core of QKD is the security, and there have been many security analysis methods proposed to investigate the security of different CV-QKD protocols~\cite{Diamanti_entropy_2015}. For the convenience of the security analysis, the eavesdropper's ability is usually restricted to three different levels, namely individual attacks, collective attacks, and coherent attacks. Individual attacks and collective attacks are, to some extent, to restrict the eavesdropper's (Eve's) attack ability, so that the exchanged state between Alice (sender) and Bob (receiver) can be treated as an
identical and independently distributed (i.i.d.) state, i.e., ${\rho _{{A^N}{B^N}}} = \sigma _{AB}^{ \otimes N}$ (where $N$ is the number of exchanged signals), which can simplify the security analysis. However, a protocol is unconditionally secure only when it is secure under coherent attacks, due to the fact that coherent attacks do not limit the ability of eavesdroppers, thereby the most general attacks. In the case of coherent attacks, the exchanged states between Alice and Bob do not have the i.i.d. structure anymore; thus, the security proof is complicated.

Diverse security analysis techniques have been developed to analyze the security of different protocols under coherent attacks, typically the de Finetti theorem~\cite{Leverrier_PRL_2015, Leverrier_PRL_2017},
the post-selection technique~\cite{Christandl_PRL_2009, Leverrier_PRL_2013}, and the entropic uncertainty relation (EUR)~\cite{Furrer_PRL_2012, Furrer_PRA_2014, Nat.Commun.6.8795.2015}. Those analysis methods can also be applied to analyze the quantum random number generation protocols~\cite{Marangon_PRL_2017, Xu_QST_2019}. Different analysis methods have their advantages and disadvantages, so they are suitable for the analysis of different protocols (see~\cite{Diamanti_entropy_2015} for detailed discussions). The advantages of the EUR lies in its intuitive physical meaning (corresponding to the guessing game~\cite{Coles_RMP_2017}) and the simple estimation method. Most of the work has been done in the EUR in~\cite{Furrer_PRL_2012}, except for the finite-size effect in estimating the covariance matrix (CM). However, in practical experiments, the estimation of the CM is always achieved by limited data; thus, the finite-size effect not only affects the estimation of min-entropy, but also the estimation of leakage information.

In this work, we focus on the parameter estimation of the EUR in CV-QKD, especially on the finite-size estimation of the CM, and the modified estimation on the max-entropy. The discussion involves only the squeezed state/homodyne detection-type protocols and has no assumption on Eve's ability, namely under coherent-attack cases. Due to the influence of the finite block length of the key, the estimation of the CM is inaccurate in the case of a short block length, compared with the ideal CM estimation cases (as shown in~\cite{Furrer_PRL_2012, Furrer_PRA_2014}). We exploit the parameter estimation technique developed in~\cite{Ruppert_PRA_2014} to consider the estimation of the CM under practical block sizes. Furthermore, inspired by the double-modulation method developed in~\cite{Ruppert_PRA_2014}, we propose a double-data modulation method to estimate the parameters in the security analysis effectively, and only one modulation is needed rather than two, which simplifies the experimental structure of the double-modulation protocol. Since the exchanged state can be used for both parameter estimation and key generation, the estimation of the max-entropy is modified, and the statistical fluctuation of estimating the max-entropy disappears. The simulation result shows that the modified estimation method can, to a large extent, save the key consumption.

This paper is organized as follows. In Section~\ref{Composable_Security_and_Description_of_The_Protocol}, we review the composable security frameworks in QKD and give the description of the discussed protocol. In Section~\ref{channel_parameter_estimation}, we discuss in detail the channel parameter estimation process with finite-size. In Section~\ref{Double-data modulation}, the modified parameter estimation method is proposed with double-data modulation. The numerical simulation and discussion are give in Section~\ref{Numerical_simulation_and_discussion}, and the conclusions are drawn in Section~\ref{Conclusions}.

\section{Composable Security and Description of the Protocol}
\label{Composable_Security_and_Description_of_The_Protocol}

In this work, we investigate the CV-QKD protocol under the universal composable framework (UCF), which can be seen in~\cite{Renner_PHD_2006, Muller_NJP_2009} for the details, and the discussion is under the coherent-attack cases. The UCF is of great importance to compose sequential rounds of a protocol, and even if some of the rounds are imperfect and deviate from the ideal model, the UCF can well describe their defects. A general QKD protocol can always be divided into different parts; thus, one of the benefits of UCFs is that even if part of the protocol is imperfect, this imperfection can still be applied to subsequent analysis of the rest part of the protocol to obtain the final non-ideal key. Another advantage of UCFs is that the final imperfect key generated from a QKD system can be well quantified as $\varepsilon$-secure and then can be applied to other classical communication tasks, such as the one-time pad~scenario.

To illustrate the composable security of QKD, we first use $s_A$ to denote Alice's key and use $s_B$ to denote Bob's key. In the ideal case, the keys should be correct, secret, and robust. Correctness means, for each round of the protocol, the keys of Alice and Bob are always the same, namely $s_A=s_B=S$. Secrecy means the key is independent of the third part and only known to Alice and Bob themselves. Robustness requires that, in every round of the protocol, Alice and Bob can always generate a non-empty key, namely $S \ne \bot$. If a QKD protocol can satisfy correctness, secrecy, and robustness, the protocol then can be called perfectly secure. We denote by ${\left\{ {\left| s \right\rangle } \right\}_{s \in S}}$ the orthogonal bases of the key, by ${\rho _E}$ Eve's auxiliary quantum systems, and by ${p_ \bot }$ the probability of generating an empty key set. The perfectly secure classical-quantum (cq) state between the key $S$ and the environment $E$ can be shown as follows,
\begin{equation}\label{}
\rho _{sE}^{perfect} = \left( {1-{p_ \bot }} \right)\sum\limits_{s \in S} {\frac{1}{{\left| S \right|}}\left| s \right\rangle \left\langle s \right| \otimes \rho _E^s} + {p_ \bot }\left| \bot \right\rangle \left\langle \bot \right| \otimes \rho _E^ \bot.
\end{equation}

Nevertheless, a protocol is always imperfect with practical issues, resulting in the security deviating from the ideal model. Therefore, the $\varepsilon$-security can be used to describe the practical security with imperfect features. We denote by ${\varepsilon _c}, {\varepsilon _r}, {\varepsilon _s}$ the smoothness parameters of practical correctness, robustness, and secrecy, respectively. ${\varepsilon _c}$-correctness requires that
the key in Alice and Bob's sides be different only with very small probability ${\varepsilon _c}$, namely $\Pr \left( {{s_A} \ne {s_B}} \right) \le {\varepsilon _c}$. ${\varepsilon _r}$-robustness requires that the set of the keys is empty only with a small probability, given by $\Pr \left( {S = \bot } \right) \le {\varepsilon _r}$. ${\varepsilon _s}$-secrecy can be treated as the distance between the practical security and the perfect security, in terms of the trace distance, given by $\frac{1}{2}{\left\| {{\rho _{sE}}-\rho _{sE}^{perfect}} \right\|_1} \le {\varepsilon _s}$. In summary, if a QKD protocol can contain ${\varepsilon _c}$-correctness, ${\varepsilon _r}$-robustness, and ${\varepsilon _s}$-secrecy, then the protocol can be called ${\varepsilon}$-secure, with $\varepsilon = {\varepsilon _c} + {\varepsilon _r} + {\varepsilon _s}$.

\begin{figure}[h]
	\centering
	\includegraphics[width= 0.46\textwidth]{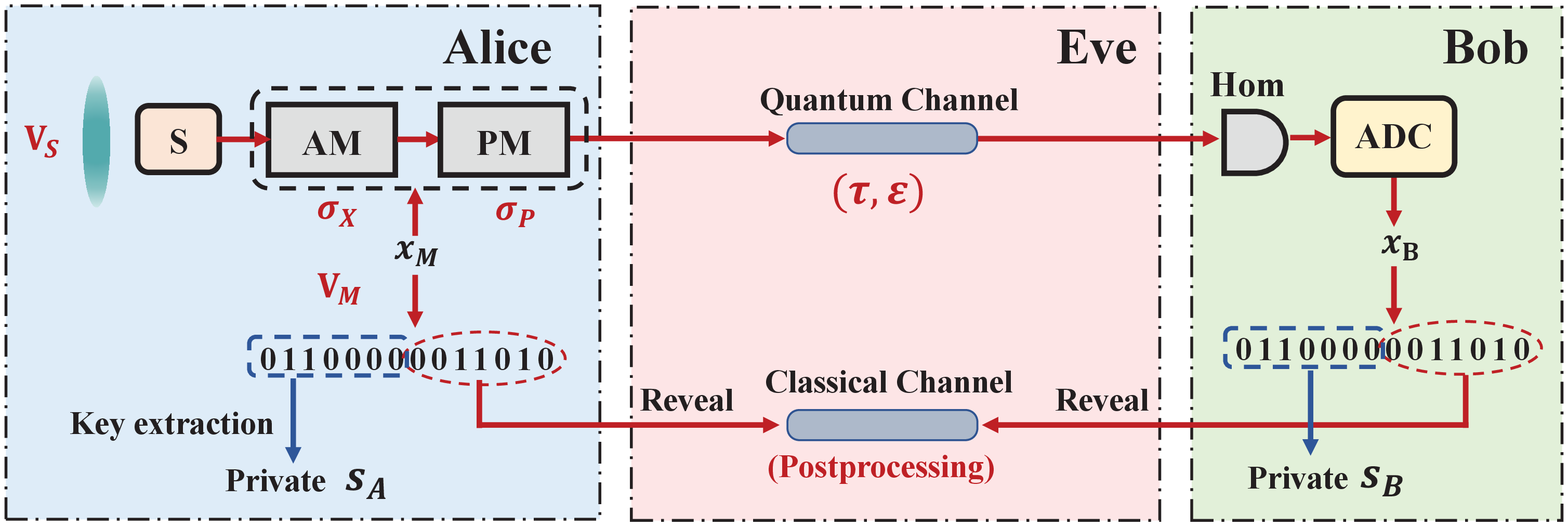}
	\caption{Prepare-and-measure (PM) scheme of continuous-variable (CV)-quantum key distribution (QKD) using squeezed states. Source: squeezed-state source with squeezed variance $V_{S}$; Mod: modulators containing amplitude and phase quadrature modulators with total modulation variance $V_M$; Hom: homodyne detection; $x_M$: Gaussian modulation data on Alice's side; $x_B$: measurement results on Bob's side; Quantum channel: channel for the transmission of quantum states, with the transmittance $\tau$ and the excess noise $\varepsilon$; Classical channel: channel for the transmission of classical data during the post-processing procedure.}
	\label{PM_scheme}
\end{figure}

Let us start with the execution of the prepare-and-measure (PM) version of the squeezed-states protocol. The protocol can be divided into sequential parts, as shown in Figure~\ref{PM_scheme}, which can be described by the following steps:

\begin{enumerate}
	\item \textbf{State preparation}: Alice holds the squeezed states with squeezed variance $V_S$ before the protocol begins, where ${V_S} \in \left( {{\rm{0,1}}} \right]$. In every run of the protocol, Alice uses Gaussian random numbers $x_M$ to encode the displacement of quadratures by using modulators (generally containing amplitude and phase modulators), and the total modulation variance is denoted by $V_M$.
	\item \textbf{State transmission}: Alice sends the modulated state in the quantum channel, which is treated as a totally untrusted channel and controlled by Eve.
	\item \textbf{State measurement}: Bob receives the quantum state and randomly measures $x$ or $p$ quadrature by an ideal homodyne detector. Resulting from the fact that the practical measurement phase is always discrete, the ideal measurement outcomes should be discretized by the analogue-to-digital converter (ADC). The final discretized results are denoted by ${x_B}$.
	\item \textbf{Parameter estimation}: Alice and Bob repeat the above steps many times until they have enough raw data (e.g., $N$). Then, Alice or Bob reveals some of the raw data (with length $m$) through the classical channel to estimate the key parameters of the channel, especially the data distance $d_0$ between Alice's and Bob's data, the transmittance $\tau$, and the excess noise $\varepsilon$. See Section~\ref{channel_parameter_estimation} for a detailed explanation of the parameter estimation step.
	\item \textbf{Error correction}: According to the estimation parameters $\tau$ and $\varepsilon$, the communication parts estimate the leakage information ${\ell _{EC}}$ during the error correction phase and choose an appropriate classical error reconciliation
	algorithm, e.g., low-density-parity-check (LDPC) code, to correct Alice's error (in reverse reconciliation cases) or Bob's error (in direct reconciliation cases).
	\item \textbf{Privacy amplification}: Alice and Bob randomly choose a universal$_2$ hash function~\cite{Carter_JCSS_1979} and apply it to their respective keys to get the final private keys $s_A$ and $s_B$ with length $\ell$, which are only known to themselves.
\end{enumerate}

According to the UCF, one can write the upper bound of the final key length ${\ell _{low}}$, even if the above steps are not ideal, given by~\cite{Renner_PHD_2006}: 
\begin{equation}\label{key_length}
{\ell _{low}} = H_{\min }^\varepsilon \left( {{x_B}|E} \right)-{\ell _{EC}}-{\log _2}\frac{1}{{\varepsilon _1^2{\varepsilon _c}}} + 2,
\end{equation}
where $H_{\min }^\varepsilon \left( {{x_B}|E} \right)$ is the smooth min-entropy
of $x_B$ conditioned
on the information Eve may hold, with smoothing parameter $\varepsilon$, and $\varepsilon_1$ is the smoothness of the physical part of the protocol.

\section{Channel Parameter Estimation with Finite-Size}
\label{channel_parameter_estimation}
There are roughly two parameters that need to be bounded in the protocol. One is the smooth min-entropy $H_{\min }^\varepsilon \left( {{x_B}|E} \right)$, and the other is the leakage information ${\ell _{EC}}$. We separately discuss the estimation of the two parameters in two parts.
\subsection{Estimation of Smooth Min-Entropy}
There are different ways to estimate the min-entropy under coherent attacks. For instance, the de Finetti theorem~\cite{Leverrier_PRL_2015, Leverrier_PRL_2017}, which can reduce the analysis from the coherent attack case to the collective attack case, has been successfully used to prove the security of CV-QKD protocols with the source of coherent states~\cite{Phys.Rev.A.97.052327.2018, Ghorai_PRA_2019}. The EUR has also been exploited to
prove the security of squeezed-state-type protocols~\cite{Furrer_PRL_2012, Furrer_PRA_2014, Chen_PRA_2018}. In this work, we focus on using the uncertainty relation to bound the min-entropy of the key. 

In practical experiments, $x_M$ and $x_B$ are always discretized. We denote $\alpha$ as the maximum discretization range of the sampling interval and denote $\delta$ as the discrete precision of the measurement, which satisfy ${{{\rm{2}}\alpha } \mathord{\left/{\vphantom {{{\rm{2}}\alpha } \delta }} \right.\kern-\nulldelimiterspace} \delta }{\rm{ = }}{{\rm{2}}^L} \in \mathbb{N}$, where $L$ is the number of discrete bits. Therefore, the measurement result will fall into different intervals, namely,
\begin{equation}\label{}
\left( {-\infty ,-\alpha } \right],...\left( {-\alpha + \left( {k-1} \right)\delta ,-\alpha + k\delta } \right],...,\left( {\alpha , + \infty } \right),
\end{equation} 
where $k = \left\{ {1,2,...,{{2\alpha } \mathord{\left/{\vphantom {{2\alpha } \delta }} \right.\kern-\nulldelimiterspace} \delta }} \right\}$. One can bound the smooth min-entropy of the discretized data $x_B$ conditioned on Eve's information $H_{\min }^\varepsilon \left( {{x_B}|E} \right)$ according to the CV version of EUR, given by:
\begin{equation}\label{EUR_BOUND}
H_{\min }^\varepsilon \left( {{x_B}|E} \right) \ge-n\log c\left( \delta \right)-H_{\max }^{\varepsilon '}\left( {{x_M}|{x_B}} \right),
\end{equation}
where $c$ quantifies the maximum overlap of the two measurements, namely $c = \mathop {\max }\limits_{x,z} {\left| {\left\langle {{{\mathbb{X}^x}}}\mathrel{\left | {\vphantom {{{\mathbb{X}^x}} {{\mathbb{Z}^z}}}}
			\right.\kern-\nulldelimiterspace}{{{\mathbb{Z}^z}}} \right\rangle }\right|^2}$ and $\mathbb{X}$ and $\mathbb{Z}$ are mutually unbiased bases; hence, $c\left( \delta \right)$ is the overlap between discrete quadrature measurements related to the interval length $\delta$, which reads:
\begin{equation}\label{}
c\left( \delta \right) = \frac{1}{{2\pi }}{\delta ^2}S_0^{\left( 1 \right)}{\left( {1,\frac{{{\delta ^2}}}{4}} \right)^2},
\end{equation}
where $S_0^{\left( 1 \right)}(.)$ is the zeroth radial prolate spheroidal wave function of the first kind~
\cite{Kiukas_JMP_2010} and $S_0^{\left( 1 \right)}{\left( {1,\frac{{{\delta ^2}}}{4}} \right)^2}$ is approximately one if $\delta$ is small. The term $H_{\max }^{\varepsilon '}\left( {{x_M}|{x_B}} \right)$ in Equation~\ref{EUR_BOUND} denotes the max-entropy between Alice's and Bob's data, with smoothing parameter $\varepsilon ' = {{{\varepsilon _s}} \mathord{\left/{\vphantom {{{\varepsilon _s}} {4{p_{pass}}}}} \right.\kern-\nulldelimiterspace} {4{p_{pass}}}}-{{2\sqrt {2\left[ {1-{{\left( {1-{p_\alpha }} \right)}^n}} \right]} } \mathord{\left/{\vphantom {{2\sqrt {2\left[ {1-{{\left( {1-{p_\alpha }} \right)}^n}} \right]} } {\sqrt {{p_{pass}}} }}} \right.\kern-\nulldelimiterspace} {\sqrt {{p_{pass}}} }}$, where $p_\alpha$ is the probability that the measurement is outside of the
detection range.

According to Equation~\ref{EUR_BOUND}, in order to give a lower bound of the min-entropy, one should estimate the upper bound of the max-entropy using some of the raw keys during the parameter estimation phase. First, the average distance, which quantifies the correlation between Alice's and Bob's data, should be estimated, given by:
\begin{equation}\label{data_correlation}
d\left( {x_M^{PE},x_B^{PE}} \right) = \frac{1}{m}\sum\limits_{i = 1}^m {\left| {{M_i}-{B_i}} \right|},
\end{equation}
where we use ${M_i}$ to denote the $i^{\text{th}}$ modulating value and ${B_i}$ denotes the $i^{\text{th}}$ measurement result, for $i=1,2,...,m$, respectively. If the data distance $d\left({x_M^{PE},x_B^{PE}} \right)$ is smaller than a certain threshold $d_0$, the parameter estimation step passes. Then, one can bound the max-entropy according to Serfling's large deviation bound~\cite{Serfling_AS_1974}, given by:
\begin{equation}\label{}
H_{\max }^\varepsilon \left( {{x_M}|{x_B}} \right) \le n{\log _2}\gamma \left( {{d_0} + \mu } \right),
\end{equation}
where $\gamma$ is a large deviation function, which reads:
\begin{equation}\label{}
\gamma (t) = \left( {t + \sqrt {{t^2} + 1} } \right){\left[ {\frac{t}{{\sqrt {{t^2} + 1}-1}}} \right]^t},
\end{equation} 
and $\mu$ quantifies the impact of statistical fluctuations resulting from estimating ``data parameter'' $H_{\max }^\varepsilon \left( {{x_M}|{x_B}} \right)$ by ``PE
parameter'' $H_{\max }^\varepsilon \left( {x_M^{PE}|x_B^{PE}} \right)$, which reads:
\begin{equation}\label{}
\mu = \frac{{2\alpha }}{\delta }\sqrt {\frac{{N\left( {m + 1} \right)}}{{n{m^2}}}\ln \frac{1}{{\varepsilon '}}},
\end{equation}
where $N$ denotes the total number of exchanged signals and satisfies $N=n+m$.


\subsection{Ideal Estimation of Leakage Information with Infinite-Size}
To estimate the leakage information in the error correction phase, we model Eve's behavior by the entangling cloner attack model, which is the most common example of a Gaussian attack~\cite{Grosshans_QIC_2003}. We point out that the whole analysis of this paper is under the most general coherent attacks and has no restriction on Eve's ability. The model of the entangling cloner attack is only for intuitive understanding, and it is convenient to investigate the performance of the protocol, which can be used to estimate the lower bound of the key rate. Even if Eve's attack is not the entangling cloner attack, the following analysis also holds, resulting from the fact that in a practical experiment, we do not need to assume the eavesdropper's strategy in advance and only need to estimate the channel parameters by the existing data that Alice and Bob hold.

The quadrature of the quantum state sent by Alice's side is denoted by ${x_A} = {x_s} + {x_M}$. In order to obtain the correlation between Alice and Bob after passing through the channel, we assume Eve performs the entangling cloner attack, where Eve's state is modeled by a two-mode squeezed vacuum (TMSV) state ${\rho _{e{E_0}}}$ with the CM ${\gamma _{e{E_0}}}$, which reads:
\begin{equation}\label{}
{\gamma _{e{E_0}}} = \left( {\begin{array}{*{20}{c}}
	{\omega \textbf{I}}&{\sqrt {{\omega ^2}-1} \textbf{Z}}\\
	{\sqrt {{\omega ^2}-1} \textbf{Z}}&{\omega \textbf{I}}
	\end{array}} \right),
\end{equation}
where $\omega$ is the variance of the TMSV, $\textbf{I} = {\rm{diag}}\left( {1,1} \right)$, and $\textbf{Z} = {\rm{diag}}\left( {1,-1} \right)$. The channel is modeled by a beam splitter with the transmittance $\tau$, whose CM is given by:
\begin{equation}\label{}
{S_\tau }{\rm{ = }}\left( {\begin{array}{*{20}{c}}
	{\sqrt \tau \textbf{I}}&{\sqrt {1-\tau } \textbf{I}}\\
	{-\sqrt {{\rm{1-}}\tau } \textbf{I}}&{\sqrt \tau \textbf{I}}
	\end{array}} \right),
\end{equation}
and the excess noise $\varepsilon$ can be defined as $\varepsilon : = {{\left( {1-\tau } \right)\left( {\omega-1} \right)} \mathord{\left/{\vphantom {{\left( {1-\tau } \right)\left( {\omega-1} \right)} \tau }} \right.	\kern-\nulldelimiterspace} \tau }$. Thus, it is easy to deduce the quadrature on Bob's side after passing through the quantum channel, given by:
\begin{equation}\label{}
{x_B} = \sqrt \tau {x_A} + \sqrt {1-\tau } {x_0} + {x_\varepsilon } = \sqrt \tau {x_M} + {x_N},
\end{equation}
where ${x_N} = \sqrt \tau {x_s} + \sqrt {1-\tau } {x_0} + {x_\varepsilon }$. Assuming that the squeezing operation is performed for $x$ quadrature, the mutual information between Alice and Bob reads:
\begin{equation}\label{}
{I^x}\left( {A:B} \right) = \frac{1}{2}{\log _2}\frac{{{V_B}}}{{{V_{B|A}}}} = \frac{1}{2}{\log _2}\left( {1 + \frac{{\tau {\sigma _x}}}{{ {V_N}}}} \right),
\end{equation}
and ${V_N}$ has the form:
\begin{equation}\label{V_N_Variance}
{V_N} = 1 + \tau \varepsilon + \tau \left( {{V_S}-1} \right): = 1 + {V_\varepsilon } + \tau \left( {{V_S}-1} \right).
\end{equation}

When Alice and Bob perform the error correction step, they need to randomly announce part of the information through the public channel, which is also revealed to Eve. It is assumed that eavesdroppers can monitor all classical communication processes; thus, the amount of information leaked in the error correction process must be well estimated and then removed from the final keys. The leakage information ${\ell _{EC}}$ in the error correction step can be described as
\begin{equation}\label{}
{\ell _{EC}^{DR}}{\rm{ = }}H({x_M})-\beta {I^x}\left( {A:B} \right),
\end{equation}
in the direct reconciliation (DR) case and:
\begin{equation}\label{}
\ell _{EC}^{RR}{\rm{ = }}H({x_B})-\beta {I^x}\left( {A:B} \right),
\end{equation}
in the reverse reconciliation (RR) case, where $\beta$ is the reconciliation efficiency.

\subsection{Practical Estimation of Leakage Information with Finite-Size}

In the previous works, the estimator of the leakage information ${\hat \ell _{EC}}$ was treated as an asymptotic parameter, which is independent of the total key length. However in practice, the estimation of ${\hat \ell _{EC}}$ cannot be accurate especially when the key length is not large, further affecting the performance of the error correction. To take finite-size effects into consideration, the estimator ${\hat \ell _{EC}}$ under a practical block length needs to be estimated. We adapt the estimation method shown in~\cite{Ruppert_PRA_2014} to analyze the characteristics of the channel. Here, we only give the main results of the previous work, and the detailed derivation can be seen in~\cite{Ruppert_PRA_2014}. In the practical experiment, the data on Alice's side is actually the modulated data $x_M$; thus, the key of parameter estimation is to estimate the CM ${\gamma _{MB}}$, namely ${\gamma _{MB}} = \left[ {{V_M}\textbf{I},{c_{MB}}\textbf{Z};{c_{MB}}\textbf{Z},{V_B}\textbf{I}} \right]$. The relation of $x_M$ and $x_B$ (Alice's and Bob's data) has the form of ${x_B} = \sqrt \tau {x_M} + {x_N}$, where ${x_N}$ is the aggregated noise with zero mean, and the variance is shown in Equation~\ref{V_N_Variance}. The covariance of ${x_M}$ and ${x_B}$ is:
\begin{equation}\label{cov_origin}
Cov\left( {{x_M},{x_B}} \right) = \sqrt \tau {V_M} = :{c_{MB}}.
\end{equation}

For obtaining the estimator of covariance ${\hat c_{MB}}$, we also use ${M_i}$ denoting the $i^{\text{th}}$ modulating value and ${B_i}$ denoting the $i^{\text{th}}$ measurement result, for $i=1,2,...,m$, respectively. According to the maximum likelihood estimation, we can get:
\begin{equation}\label{}
{\hat c_{MB}} = \frac{1}{m}\sum\limits_{i = 1}^m {{M_i}{B_i}}.
\end{equation}
and it is easy to compute the expectation value $\mathbb{E} \left[ {{{\hat c}_{MB}}} \right]$ and the variance $\mathbb{V} \left[ {{{\hat c}_{MB}}} \right]$ by assuming $M_i$ and $B_i$ are two independent Gaussian variables with zero mean values, which read:
\begin{equation}\label{}
\mathbb{E}\left[ {{{\hat c}_{MB}}} \right] = {c_{MB}},
\end{equation}
\begin{equation}\label{}
\mathbb{V}\left[ {{{\hat c}_{MB}}} \right] = \frac{{\tau V_M^2}}{m}\left( {2 + \frac{{{V_N}}}{{\tau {V_M}}}} \right).
\end{equation}

According to Equation~\ref{cov_origin}, we can get the estimator $\hat \tau$ of $\tau$, which reads:
\begin{equation}\label{}
\hat \tau = \frac{{\hat c_{MB}^2}}{{V_M^2}} = \frac{{\mathbb{V}\left[ {{{\hat c}_{MB}}} \right]}}{{V_M^2}}{\left( {\frac{{{{\hat c}_{MB}}}}{{\sqrt {\mathbb{V}\left[ {{{\hat c}_{MB}}} \right]} }}} \right)^2},
\end{equation}
where ${\left( {\frac{{{{\hat c}_{MB}}}}{{\sqrt {\mathbb{V}\left[ {{{\hat c}_{MB}}} \right]} }}} \right)^2}$ follows the ${\chi ^2}$-distribution, namely, 
\begin{equation}\label{}
{\left( {\frac{{{{\hat c}_{MB}}}}{{\sqrt {\mathbb{V}\left[ {{{\hat c}_{MB}}} \right]} }}} \right)^2} \sim {\chi ^2}\left( {1,\frac{{\hat c_{MB}^2}}{{\mathbb{V}\left[ {{{\hat c}_{MB}}} \right]}}} \right).
\end{equation}

Then, we can calculate the expectation value of $\hat \tau$, which reads:
\begin{align}
\mathbb{E}\left( {\hat \tau } \right) = \tau + O\left( {{1 \mathord{\left/{\vphantom {1 m}} \right.	\kern-\nulldelimiterspace} m}} \right),
\end{align}
and the variance is given by:
\begin{align}
\mathbb{V}\left( {\hat \tau } \right) = \frac{{4{\tau ^2}}}{m}\left( {2 + \frac{{{V_N}}}{{\tau {V_M}}}} \right) + O\left( {{1 \mathord{\left/
			{\vphantom {1 {{m^2}}}} \right.
			\kern-\nulldelimiterspace} {{m^2}}}} \right).
\end{align}

For $m \gg 1$, which is practical in experiments, the term $O\left( {{1 \mathord{\left/{\vphantom {1 {{m^2}}}} \right.\kern-\nulldelimiterspace} {{m^2}}}} \right)$ can be negligible due to the order ${{1 \mathord{\left/{\vphantom {1 {{m^2}}}} \right.\kern-\nulldelimiterspace} {{m^2}}}}$ being small. Thus, we define new variance of $\hat \tau$ under a practical block length, which reads:
\begin{equation}\label{}
\sigma _{\hat \tau }^{\rm{2}} = \frac{{4{\tau ^2}}}{m}\left( {2 + \frac{{{V_N}}}{{\tau {V_M}}}} \right),
\end{equation}
so that the confidence interval of estimating $\tau$ can be well quantified. 

In order to estimate the upper bound of the leakage information $\ell _{EC}^{up}$, one should give the lower bound of the transmittance $\tau$. For practical purposes, we set the failure probability of the parameter estimation to ${\varepsilon _{PE}} = {10^{-10}}$, which corresponds to the confidence interval of $6.5{\sigma _{\hat \tau }}$, and one can estimate the lower bound of ${\hat \tau ^{low}}$, given by:
\begin{equation}\label{}
{\hat \tau ^{low}} = \mathbb{E}\left( {{\tau ^{low}}} \right): = \hat \tau-6.5{\sigma _{\hat \tau }}.
\end{equation}

According to:
\begin{equation}\label{evolution_single_modulation}
{x_B} = \sqrt \tau \left( {{x_M} + {x_S}} \right) + \sqrt {1-\tau } {x_0} + {x_\varepsilon } = \sqrt \tau {x_M} + {x_N},
\end{equation}
the estimator of ${V_\varepsilon }$ can also be calculated by the maximum likelihood estimation with the following~form:
\begin{equation}\label{}
{\hat V_\varepsilon } = \frac{1}{m}\sum\limits_{i = 1}^m {{{\left( {{B_i}-\sqrt {\hat \tau } {M_i}} \right)}^2}} + \hat \tau \left( {1-{V_S}} \right)-1.
\end{equation}

In the case of $m \gg 1$, the estimator $\hat \tau$ converges rapidly to the actual value $\tau$ as $m$ increases, owing to the variance of $\hat \tau$ being negligible. Thus, here, we use $\tau$ to replace $\hat \tau$ to simplify the estimation process. Noticing that the term $\frac{1}{m}\sum\limits_{i = 1}^m {{{\left( {\frac{{{B_i}-\sqrt \tau {M_i}}}{{\sqrt {{V_N}} }}} \right)}^2}}$ also follows the ${\chi ^2}$-distribution with the expectation value $\mathbb{E}\left( {\frac{1}{m}\sum\limits_{i = 1}^m {{{\left( {\frac{{{B_i}-\sqrt \tau {M_i}}}{{\sqrt {{V_N}} }}} \right)}^2}} } \right) = m$ and variance $\mathbb{V}\left({\frac{1}{m}\sum\limits_{i = 1}^m {{{\left( {\frac{{{B_i}-\sqrt \tau {M_i}}}{{\sqrt {{V_N}} }}} \right)}^2}} } \right) = 2m$, respectively, resulting from ${{B_i}-\sqrt \tau {M_i}}$ being Gaussian distributed with variance $V_N$, therefore, one can get the following approximation when $m$ is large:
\begin{equation}\label{}
\sum\limits_{i = 1}^m {{{\left( {{B_i}-\sqrt \tau {M_i}} \right)}^2}} \approx {V_N} \cdot \sum\limits_{i = 1}^m {{{\left( {\frac{{{B_i}-\sqrt \tau {M_i}}}{{\sqrt {{V_N}} }}} \right)}^2}}.
\end{equation}

The expectation value of ${\hat V_\varepsilon }$ can be obtained, which reads:
\begin{align}
\mathbb{E}\left( {{{\hat V}_\varepsilon }} \right)&\approx \frac{1}{m}{V_N} \cdot \mathbb{E}\left( {\sum\limits_{i = 1}^m {{{\left( {\frac{{{B_i}-\sqrt \tau {M_i}}}{{\sqrt {{V_N}} }}} \right)}^2}} } \right)   \notag\\
&+ \tau \left( {1-{V_S}} \right)-1 = {V_\varepsilon },
\end{align}
and the variance of ${\hat V_\varepsilon }$ can also be calculated, given by:
\begin{align}
\mathbb{V}\left( {{{\hat V}_\varepsilon }} \right) \approx \frac{{\rm{2}}}{m}V_N^2 + \sigma _{\hat \tau }^{\rm{2}}{\left( {1-{V_S}} \right)^{\rm{2}}}:=\sigma _{{{\hat V}_\varepsilon }}^{\rm{2}}.
\end{align}

The upper bound of the variance of excess noise can be given, also considering the failure probability of the parameter estimation to ${\varepsilon _{PE}} = {10^{-10}}$, which is:
\begin{equation}\label{}
\hat V_\varepsilon ^{up} = \mathbb{E}\left( {V_\varepsilon ^{up}} \right): = {\hat V_\varepsilon } + 6.5{\sigma _{{{\hat V}_\varepsilon }}}.
\end{equation}

\section{Double-Data Modulation Method and the Modified Estimation Process}
\label{Double-data modulation}

Inspired by the double-modulation method developed in~\cite{Ruppert_PRA_2014}, we find that this estimation method is also useful in the parameter estimation of the EUR analysis method.

Here, we slightly modify the double-modulation method by pre-generating two sets of Gaussian random numbers, namely $x_{M1}$ and $x_{M2}$, with variances $V_{M1}$ and $V_{M2}$ and zero mean values, encoding quantum states by new random variable ${x_M}$, where ${x_M} = {x_{M1}} + {x_{M2}}$. In this double-data modulation method, Alice holds both data ${x_{M1}}$ and ${x_{M2}}$ in her memories and then generates data $x_M$ according to data ${x_{M1}}$ and ${x_{M2}}$. The generated data $x_M$ are used to modulate the quantum states. After Alice and Bob finish the key distribution processes, Alice reveals data ${x_{M2}}$ to perform the channel parameter estimation, and all the information about data ${x_{M1}}$ is not announced throughout the parameter estimation phase; thus, ${x_{M1}}$ can be used for the key extraction step without leaking information about the key during the parameter estimation step. The idea is very similar to that in~\cite{Ruppert_PRA_2014}, and the difference is that this double-data modulation method only needs one modulation rather than two, since we perform the pre-processing of two independent random variables, which simplifies the experimental setup of the double-modulation method.

Since all the exchanged signals can be used for both parameter estimation and key extraction, the estimation of the max-entropy needs to be modified. Recalling that in Section~\ref{channel_parameter_estimation}, the key point of estimating the max-entropy is to quantify the data distance $d\left( {x_M^{total},x_B^{total}} \right)$. However, in traditional EUR method, not all the data can be used for the parameter estimation, and only part of the data (parameter estimation data) can be used to estimate the total data distance, resulting in the statistical fluctuation of the estimating distance, thereby $d\left( {x_M^{total},x_B^{total}} \right)$ is approximately replaced by $d\left( {x_M^{PE},x_B^{PE}} \right) + \mu$, where the first term is the distance between the parameter estimation data and the second term is the statistical fluctuation of estimating the total data distance by using the parameter estimation data. In the double-data modulation protocol, we modify the $L_1$ distance between the key-extraction data $x_{M1}$
and Bob's data ${x_B}$ by exploiting the absolute value inequality, given by: 
\begin{align}
d\left( {{x_{M1}},{x_B}} \right) &= \frac{1}{N}\sum\limits_N {\left| {x_B^i-x_{M1}^i} \right|} \notag \\ 
&\le \frac{1}{N}\sum\limits_N {\left| {x_B^i-x_{M2}^i} \right|} + \frac{1}{N}\sum\limits_N {\left| {x_{M2}^i-x_{M1}^i} \right|}   \notag \\ 
&= d\left( {{x_{M2}},{x_B}} \right) + d\left( {{x_{M1}},{x_{M2}}} \right),
\label{L1_distance}
\end{align}
where $d\left( {{x_{M2}},{x_B}} \right)$ denotes the $L_1$ distance between data $x_{M2}$ and ${x_B}$, which can be estimated after Alice reveals data $x_{M2}$, and $d\left( {{x_{M1}},{x_{M2}}} \right)$ denotes the $L_1$ distance between data $x_{M1}$ and $x_{M2}$, which can be calculated on Alice's side locally. Here, we replace the number of parameter estimation signals $m$ by $N$ since all the exchanged signals are used in this step. Therefore, the max-entropy can be bounded after modifying the parameter estimation step, which reads:
\begin{equation}\label{}
H_{\max }^\varepsilon \left( {{x_{M1}}|{x_B}} \right) \le N{\log _2}\left( {d\left( {{x_{M2}},{x_B}} \right) + d\left( {{x_{M1}},{x_{M2}}} \right)} \right).
\end{equation}

Due to the fact that all the states are exploited to perform parameter estimation, the statistical fluctuation of estimating $L_1$ distance disappears, which reduces the finite-size effect on estimating the max-entropy, especially in the short block size regime, where the statistical fluctuation cannot be~negligible.

The remaining task is to estimate the confidence intervals of the channel parameters by using data $x_{M2}$ and $x_B$, which is the standard estimation method shown in~\cite{Ruppert_PRA_2014}. The quadrature of the received states on Bob's side can be rewritten in the following form after using the double-data modulation method,
\begin{align}\label{evolution_double_modulation}
{x_B} &= \sqrt \tau \left( {{x_M} + {x_S}} \right) + \sqrt {1-\tau } {x_0} + {x_\varepsilon }  \notag \\ 
&= \sqrt \tau {x_{M2}} + {x_N^ *},
\end{align}
where $x_N^ * = \sqrt \tau \left( {{x_s} + {x_1}} \right) + \sqrt {1-\tau } {x_0} + {x_\varepsilon }$ is the aggregated noise when we use $x_{M2}$ to perform the parameter estimation, with variance $V_N^ * = \tau \left( {{x_s} + {x_1}-1} \right) + {\rm{1}} + {V_\varepsilon }$.

After comparing Equation~\ref{evolution_double_modulation} with Equation~\ref{evolution_single_modulation}, it is easy to obtain the variances of the estimators $\hat \tau$ and ${\hat V_\varepsilon }$ by replacing $V_M$ with $V_{M2}$, $V_{N}$ with $V_N^ *$, and $m$ with $N$, which are given by:
\begin{align}
\sigma _{{{\hat \tau }^ * }}^2 &= \frac{{4{\tau ^2}}}{N}\left( {2 + \frac{{{V_N^*}}}{{\tau {V_{M2}}}}} \right), \\
\sigma _{\hat V_\varepsilon ^ * }^2 & = \frac{{\rm{2}}}{N} {V_N^*}^2 + \sigma _{\hat \tau^* }^{\rm{2}}{\left( {1-{V_S}} \right)^{\rm{2}}}.
\end{align}

\section{Numerical Simulation and Discussion}
\label{Numerical_simulation_and_discussion}

In this section, we focus on the simulation analysis of the protocol with the finite-size effect, containing the
comparison of the protocol's performances between ideal and practical estimations of the CM and the comparison between standard estimation method and the modified double-data modulation method. The simulation assumes that Eve's attack is the entangling cloner attack. We stress again that this attack model does not affect the security of the protocol and is just for the convenience of the simulation. In practice, we do not need to assume the attack model in advance and only need to estimate the correlation through the data in the hands of Alice and Bob. The correlation between Alice's and Bob's data can be verified according to whether the $L_1$ distance $d\left( {x_M^{PE},x_B^{PE}} \right)$ shown in Equation~\ref{data_correlation} is greater than the threshold parameter $d_0$. If the relation $d\left( {x_M^{PE},x_B^{PE}} \right) < {d_0}$ holds, we think the data between Alice and Bob are correlated. Otherwise, we abort the protocol. In order to determine whether the amount of data is sufficient for the parameter estimation, one needs to use the experimental data of Alice and Bob with a finite block size to estimate the practical parameters and to determine whether the finite-size effect is acceptable by simulation.

We point out that the analysis using the EUR does not rely on Eve's attack method in the experiment, which is due to two reasons. One reason is that the EUR security analysis method itself does not restrict Eve's ability~\cite{Furrer_PRL_2012}, which means there is no need to assume that the quantum state is a product state $\sigma _{AB}^{ \otimes N}$, like the collective-attack analysis. Another reason is that the parameter estimation does not need to assume Eve's attacking model. The estimation of max-entropy only needs to estimate the data distance $d\left( {x_M^{PE},x_B^{PE}} \right)$ by $x_M$ and $x_B$. The estimation of ${\ell _{EC}}$ needs the variance of the measured data and the signal-to-noise ratio after transmission, which can be obtained from the statistical CM directly. Using the entangling cloner attack model to model Eve's behavior just aims at getting the lower bound of the transmittance $\tau$ and the upper bound of the excess noise $\varepsilon$, and then, the lower bound of the key rate can be calculated.

In the following discussion, we consider the squeezed vacuum states with a squeezing level of 13.1 dB and an anti-squeezing level of 25.8 dB, which has experimentally been achieved at 1550 nm with today's technology~\cite{Schonbeck_OL_2018}. We set the reconciliation efficiency $\beta$ to $95\%$, which is also easily achievable with CV-QKD's post-processing method~\cite{Wang_QIC_2017, Wang_SR_2018}. The excess noise is chosen as $\varepsilon = 0.01$, and the security parameters are chosen as ${\varepsilon _c} = {\varepsilon _s} = {10^{-9}}$.

\begin{figure}[h]
	\centering
	\includegraphics[width= 0.46\textwidth]{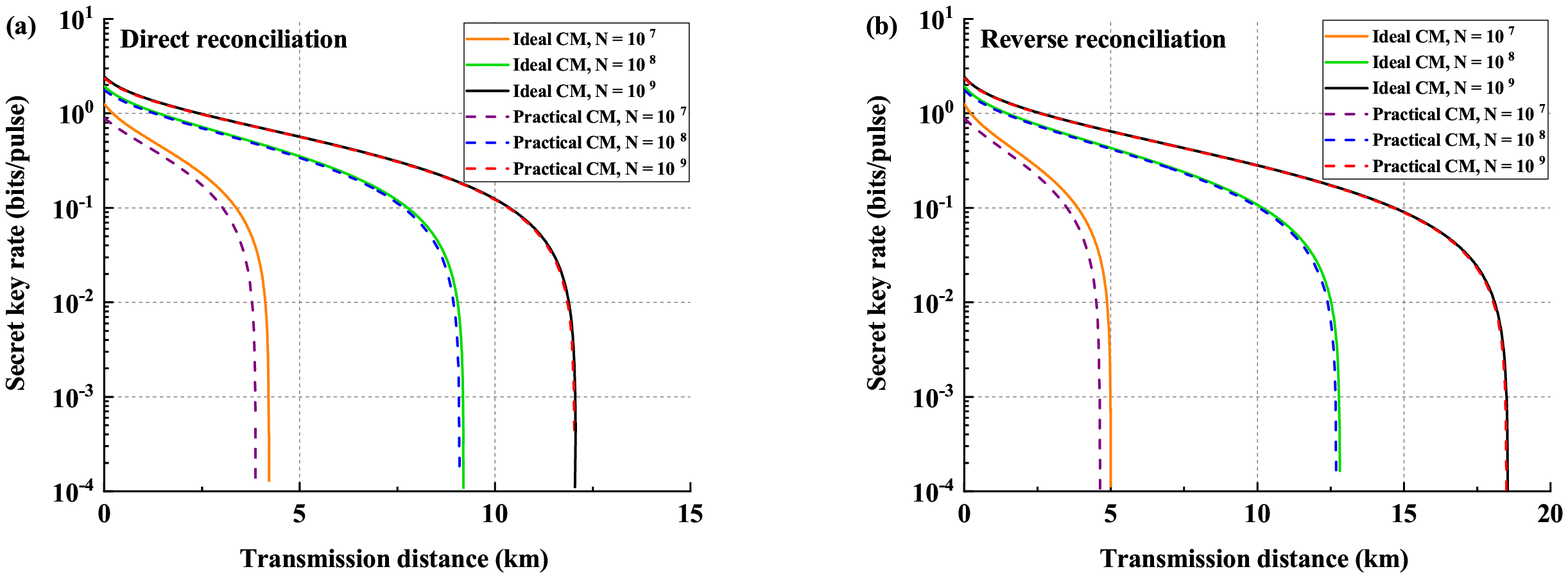}
	\caption{Comparison of performances between the previous key rates and the modified results under different block lengths, namely, $10^7$, $10^8$, and $10^9$. (a) shows the direct reconciliation (DR) cases, and (b) shows the reverse reconciliation (RR) cases. The solid lines are the performances under the ideal covariance matrix (CM) estimation, and the dashed lines are the performances under practical CM estimation considering finite-size. The reconciliation efficiency $\beta$ is under a practical value of $95\%$, and the excess noise is chosen as $\varepsilon {\rm{ = 0}}{\rm{.01}}$. We set the security parameters ${\varepsilon _c} = {\varepsilon _s} = {10^{-9}}$ and the detection range to $\alpha = 61.6$.}
	\label{key_distance}
\end{figure}

In Figure~\ref{key_distance}, we plot the key rate as a function of the transmission distance, expressed in terms of km. The lower bound of the key length is given by Equation~\ref{key_length}, and the secret key rate is calculated by ${{{\ell _{low}}} \mathord{\left/{\vphantom {{{\ell _{low}}} N}} \right. \kern-\nulldelimiterspace} N}$. The left panel and the right panel are the performances under the DR and RR cases, respectively. We give the comparison between the ideal CM estimation and the practical CM estimation with different practical block sizes, namely $10^7$, $10^8$, and $10^9$. The solid lines are the protocol under ideal CM estimation, and the dashed lines are the performances under practical CM estimation. We can find that the finite-size effect of estimating the CM will slightly influence the final key rates, and the larger the block size, the smaller the impact. For a practical block size of the order of $10^9$, there is almost no influence on the secret key rate.

\begin{figure}[h]
	\centering
	\includegraphics[width= 0.46\textwidth]{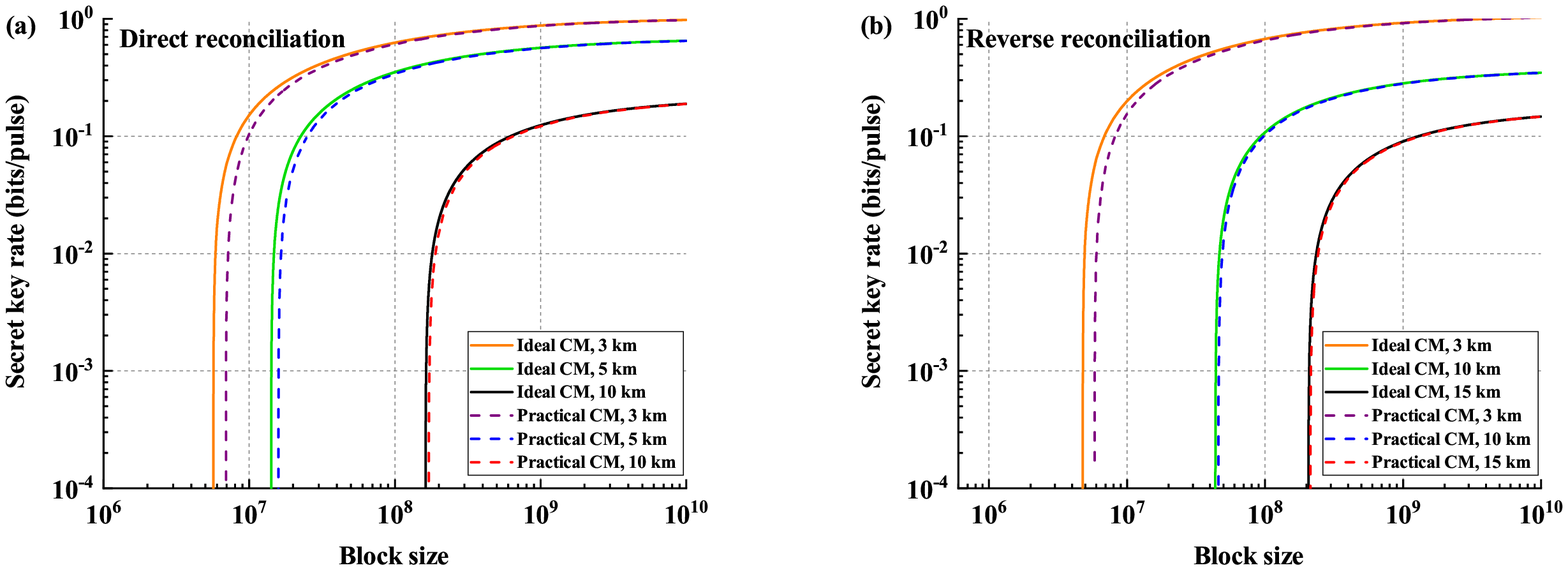}
	\caption{Comparison of performances between the previous key rates and the modified results under different transmission distances. (a) shows the direct reconciliation cases, and (b) shows the reverse reconciliation cases. The solid lines are the performances under ideal CM estimation, and the dashed lines are the performances under practical CM estimation considering finite-size. The parameters are chosen as in Figure~\ref{key_distance}.}
	\label{key_N}
\end{figure}

In Figure~\ref{key_N}, we plot the key rate of the protocol as a function of the block size and compare the performances under different transmission distances. In the DR case (left panel), the performances under transmission distances of 3 km, 5 km, and 10 km are illustrated, while the key rates under transmission distances of 3 km, 10 km, and 15 km are plotted in the RR case (right panel), respectively. We can see that the block length of the order of $10^7-10^9$ is sufficient for the protocol under the composable security analysis, achieving rates over $10^{-1}$ bits per channel use for transmission distances of about 10 km in DR and 15 km in RR, respectively. The results also show that, in the case of short transmission distance, the limited block length has a small impact on the performance of the protocol, which will be weakened with the increase of the block length. Moreover, in the case of relatively long transmission distance (approximately more than 10 km), the estimation of leakage information with finite-size has little effect on the final key since the case of long transmission distance requires a larger block size for the error correction.

\begin{figure}[h]
	\centering
	\includegraphics[width= 0.46\textwidth]{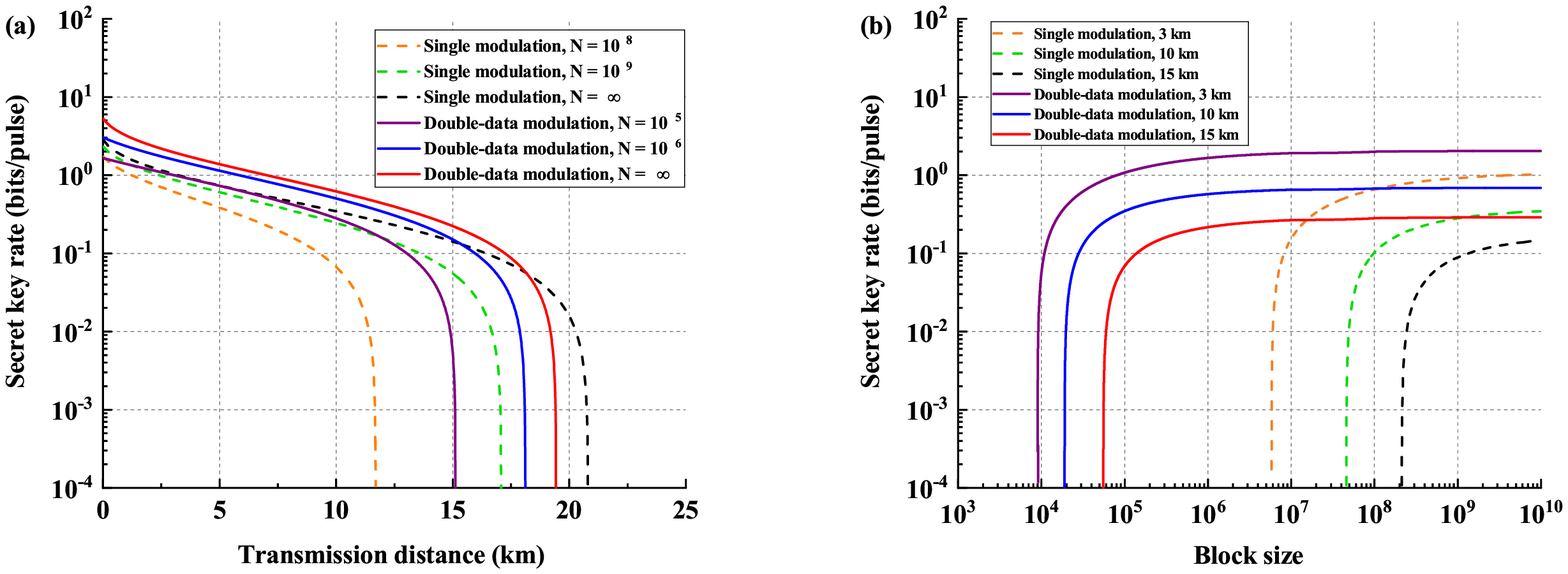}
	\caption{Comparison of the performances between the standard estimation method and the modified double-data modulation method under the reverse reconciliation case. (a) shows the performances of two scenarios under different block sizes, while (b) shows the protocol's performances under different transmission distances. The dashed lines are the performances using the standard estimation method, and the solid lines are the performances using double-data modulation~method.}
	\label{key_double_modulation}
\end{figure}

The comparison of the performances between the standard estimation method and the modified double-data modulation method is shown in Figure~\ref{key_double_modulation}, where the left panel shows the performances of two scenarios under different block sizes, while the right panel shows the protocol's performances under different transmission distances. We optimize the performance of the double-data method by adopting the optimization method shown in~\cite{Ruppert_PRA_2014}. In the left panel, we plot the performances of the double-data modulation method under block sizes of $10^5$ and $10^6$ and the asymptotic case, respectively, which are shown with solid lines, while the performances of the standard estimation method are depicted with dashed lines, under block sizes of $10^8$ and $10^9$ and the asymptotic case. It can be seen that, with the help of the double-data modulation method, using less quantum states can achieve better performance than the standard estimation method in a short block-size regime, due to the fact that the data fluctuation term $\mu$ in the previous estimation method is not negligible when the block-size is not large, which makes the statistical fluctuation of the finite-size effect more significant in short key lengths. Thus, the double-data modulation method can efficiently improve the parameter estimation process when the block size is not large. We also note that since we use all the states to extract the key, leading to a high utilization of quantum states, the key rate of the modified method is higher than that of the previous method. However, the double-data modulation method cannot achieve the transmission distance as far as the single-modulation method in the asymptotic case. This is intuitive since the statistical fluctuation in the standard estimation method converges to zero with $N$ going to infinity, while there still exit some noises in estimating data distance in double-data modulation method, namely
$d\left( {{x_{M1}},{x_{M2}}} \right)$, which will compromise the transmission distance. In the right panel of Figure~\ref{key_double_modulation}, we can see that the block length of the order of $10^5-10^7$ is sufficient for the protocol to support the previous transmission distances with the block size of the order of $10^7-10^9$, which we believe, to a large extent, saves the key consumption.

\section{Conclusions}
\label{Conclusions}

In this work, we investigated the EUR used for the composable security analysis of the CV-QKD protocol and focused on the parameter estimation step, containing the finite-size effect on estimating the CM and the improvement of the parameter the estimation phase using the double-data modulation method, which were not discussed in previous works~\cite{Furrer_PRL_2012, Furrer_PRA_2014, Nat.Commun.6.8795.2015}. We believe it is necessary to study the finite-size effect on the parameter estimation in the EUR method, as well as its improvement, since in practice, only limited exchanged states can be used for the parameter estimation, making the estimation process~non-ideal.

The analysis showed that the finite-size effect of estimating the CM had a slight influence on the key rate. The larger the block size, the smaller the influence. For a practical block length of the order of 
$10^9$, the influence on the protocol's performance was almost negligible. Thus, in a practical experiment, if the amount of data is large, treating the estimators of parameters as ideal parameters will not have a great influence on the key rate. The result also showed that the parameter estimation method developed in~\cite{Ruppert_PRA_2014} was very effective at handling the finite-size analysis of the covariance matrix in EUR~analysis.

To further reduce the impact of the finite-size effect in the parameter estimation phase, we also improved the parameter estimation process by exploiting the double-data modulation method, which was inspired by L
. Ruppert, et al~\cite{Ruppert_PRA_2014}. All the quantum states can be used for both parameter estimation and key extraction, which improves the utilization of exchanged states. After modifying the estimation of the max-entropy, we found that the finite-size effect was to a large extent suppressed when the block size was not large, which saved the key consumption, while the longest transmission distances in the asymptotic case were compromised.

Our work is an improvement of previous works~\cite{Furrer_PRL_2012, Furrer_PRA_2014}. We believe that the modified estimation method is practical by using less states to perform parameter estimation.

\begin{acknowledgments}
We would like to thank Tobias Gehring for the valuable discussions. This work is supported by the National Natural Science Foundation under Grant No. 61531003, the National Science Fund for Distinguished Young Scholars of China (Grant No. 61225003), and the China Postdoctoral Science Foundation (Grant No. 2018M630116).
\end{acknowledgments}


\bibliographystyle{unsrt}

\end{document}